\documentclass[%
 reprint,
 amsmath,amssymb,
 aps,
]{revtex4-1}

\usepackage{graphicx}
\usepackage{dcolumn}
\usepackage{bm,bbm}
\usepackage{xcolor} 
\usepackage{hyperref}

\usepackage{tcolorbox}
\definecolor{mycolor}{rgb}{0.122, 0.435, 0.698}
\tcbuselibrary{theorems}
\newtcbtheorem{Definitions}{Definition}%
{colback=mycolor!5,colframe=mycolor,fonttitle=\bfseries,
left=4pt,
right=4pt,
top=5pt,
bottom=5pt}{defi}
\newtcbtheorem{Observations}{Observation}%
{colback=green!5,colframe=green!50!blue,fonttitle=\bfseries,
left=4pt,
right=4pt,
top=5pt,
bottom=5pt,
title={#2}
}{lemma}
\newtcbtheorem{Results}{Result}%
{colback=purple!5,colframe=purple!50!blue,fonttitle=\bfseries,
left=4pt,
right=4pt,
top=5pt,
bottom=5pt
}{lemma}

\begin{document}

\title{Efficient generation of high-dimensional entanglement through multi-path downconversion}
\author{Xiao-Min Hu}
\email{These two authors contributed equally to this work.}
\affiliation{CAS Key Laboratory of Quantum Information, University of Science and Technology of China, Hefei, 230026, People's Republic of China}
\affiliation{CAS Center For Excellence in Quantum Information and Quantum Physics, University of Science and Technology of China, Hefei, 230026, People's Republic of China}
\author{Wen-Bo Xing}
\email{These two authors contributed equally to this work.}
\affiliation{CAS Key Laboratory of Quantum Information, University of Science and Technology of China, Hefei, 230026, People's Republic of China}
\affiliation{CAS Center For Excellence in Quantum Information and Quantum Physics, University of Science and Technology of China, Hefei, 230026, People's Republic of China}
\author{Bi-Heng Liu}
\email{bhliu@ustc.edu.cn}
\affiliation{CAS Key Laboratory of Quantum Information, University of Science and Technology of China, Hefei, 230026, People's Republic of China}
\affiliation{CAS Center For Excellence in Quantum Information and Quantum Physics, University of Science and Technology of China, Hefei, 230026, People's Republic of China}
\author{Yun-Feng Huang}
\affiliation{CAS Key Laboratory of Quantum Information, University of Science and Technology of China, Hefei, 230026, People's Republic of China}
\affiliation{CAS Center For Excellence in Quantum Information and Quantum Physics, University of Science and Technology of China, Hefei, 230026, People's Republic of China}
\author{Chuan-Feng Li}
\email{cfli@ustc.edu.cn}
\affiliation{CAS Key Laboratory of Quantum Information, University of Science and Technology of China, Hefei, 230026, People's Republic of China}
\affiliation{CAS Center For Excellence in Quantum Information and Quantum Physics, University of Science and Technology of China, Hefei, 230026, People's Republic of China}
\author{Guang-Can Guo}
\affiliation{CAS Key Laboratory of Quantum Information, University of Science and Technology of China, Hefei, 230026, People's Republic of China}
\affiliation{CAS Center For Excellence in Quantum Information and Quantum Physics, University of Science and Technology of China, Hefei, 230026, People's Republic of China}
\author{Paul Erker}
\affiliation{Institute for Quantum Optics and Quantum Information - IQOQI Vienna, Austrian Academy of Sciences, Boltzmanngasse 3, 1090 Vienna, Austria}%
\author{Marcus Huber}
\email{marcus.huber@univie.ac.at}
\affiliation{Institute for Quantum Optics and Quantum Information - IQOQI Vienna, Austrian Academy of Sciences, Boltzmanngasse 3, 1090 Vienna, Austria}%
\begin{abstract}
High-dimensional entanglement promises to greatly enhance the performance of quantum communication and enable quantum advantages unreachable by qubit entanglement. One of the great challenges, however, is the reliable production, distribution and local certification of high-dimensional sources of entanglement. In this article, we present an optical setup capable of producing quantum states with an exceptionally high-level of scalability, control and quality, that, together with novel certification techniques, achieve the highest amount of entanglement recorded so far. We showcase entanglement in $32$-spatial dimensions with record fidelity to the maximally entangled state ($F=0.933\pm0.001$) and introduce measurement efficient schemes to certify entanglement of formation ($E_{oF}=3.728\pm0.006$). Combined with the existing multi-core fibre technology, our results will lay a solid foundation for the construction of high-dimensional quantum networks.
\end{abstract}
\maketitle

Quantum communication often relies on entanglement to outperform classical communication tasks. The most prominent task being quantum cryptography, where entanglement enables implementations that do not require trust in the device used to create and distribute the quantum states. While most experiments and applications have thus far focused on qubits, i.e. encoding two distinguishable states in single photon pairs, recent efforts in encoding more information per photon, i.e. high-dimensional encoding, have proven successful using various physical degrees of freedom \cite{Friis_2018,Cozzolino_2019b}. From a binned time-of-arrival \cite{Tiranov_2017,Martin_2017},  orbital angular momentum (OAM) \cite{Dada:2011vc,Krenn:2017hz,Bavaresco_2018}, multiple frequencies \cite{kues2017chip} or multiple paths in silicon waveguides \cite{Wang_2018,Llewellyn_2019}, more and more possible setups have entered the stage. 

It is clear that using additional degrees of freedom, enables more bits to be encoded per photon. Increasing the encoding dimension, however, brings about novel challenges that need to be overcome to unlock that potential. The first concerns the creation of high-dimensionally entangled states. For many applications, using higher-dimensional systems only makes sense if truly more entanglement can be generated by the sources. This has thus far provided a natural limit for many systems, as for instance OAM entanglement is typically very far from potentially maximal entanglement \cite{Krenn:2017hz,Bavaresco_2018}. The second challenge is the distribution, where different encodings are not equally suited for all modes of distribution. Encoding in spatial modes rules out the most commonly used single-mode fibres and has sparked a recent shift to air-core fibres or few-mode fibres for this purpose \cite{Cozzolino_2019,cao2020distribution,Valencia:2019tc}. Finally, for the certification and indeed most applications, the possible measurements are a veritable challenge. Spatial modes are often limited by the complexity of the preceding multiport \cite{Schaeff_2015,Wang_2018,Llewellyn_2019} or the resolution of the spatial light modulators (SLMs) used prior to projective measurements \cite{Bouchard_2018}. Temporal and frequency modes on the other hand are very difficult to interfere, leading to only very limited measurement capabilities or even ruling out local measurements in some cases \cite{PhysRevLett.82.2594,Olislager_2012, kues2017chip,Chen_2020,Steinlechner_2017}. Nonetheless, the prospect for distributing entanglement in highly noise environments \cite{Ecker_2019}, makes the prospect of high-dimensional encoding exciting enough to invest efforts along all of these lines.

\begin{figure*}[tbph]
\includegraphics [width= 1\textwidth]{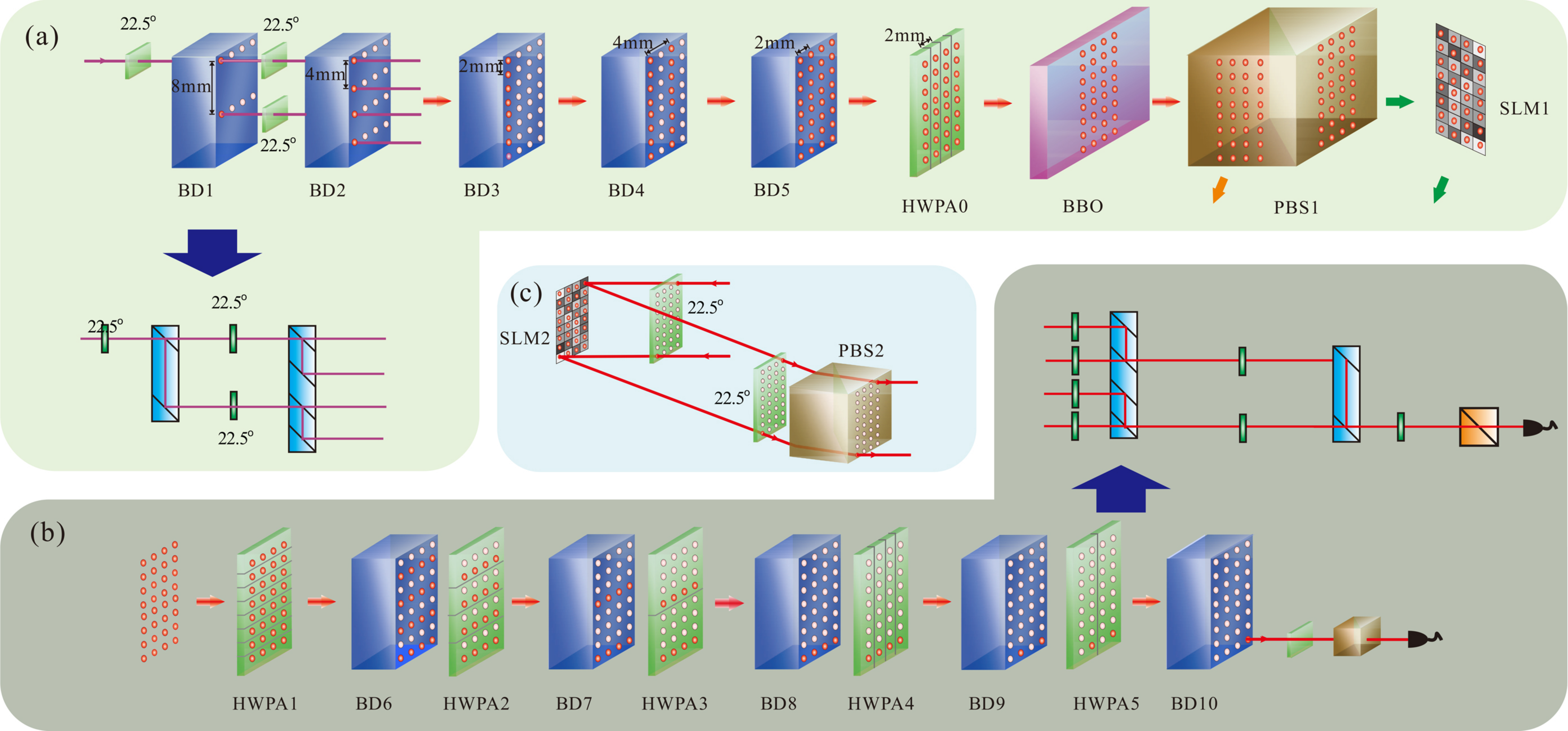}
\caption{Experimental setup. (a) The preparation of 32-dimensional entangled states. A $404~nm$ continuous wave (cw) light is divided into 32 parallel beams by half wave plates (HWPs) and beam displacers (BDs). A HWP array 0 (HWPA0) can control the polarisation of each beam and ensure that each pump beam is in the horizontal ($|H\rangle$) state. 32 beams of light effect spontaneous parametric down-conversion (SPDC) on the collinear BBO crystal ($|H\rangle_{404nm}\longrightarrow|H\rangle_{808nm}\otimes|V\rangle_{808nm}$). We encode each path as ($|0\rangle,|1\rangle,...,|31\rangle$), and can thus get the 32-dimensional entangled state $|\Phi^+\rangle=\frac{1}{\sqrt{32}} \sum_{i=0}^{31}|ii\rangle $. (b) The local measurement setup. Through the polarisation control of HWPA1-HWPA5 and BD6-BD10, an arbitrary projection measurement can be performed. In our case, we focus on the measurement of two-dimensional subspaces, which is described in detail in Fig.~\ref{fig:measurement}. (c) The beam intensity regulator. This device is implemented by spatial light modulator (SLM) which can apply an arbitrary phase to each beam. For each beam, the initial state of incident light is $|H\rangle$, HWP1 and HWP2 are set at $22.5^{\circ}$. $|H \rangle \xrightarrow{HWP1^{\circ}}\frac{1}{\sqrt{2}}(| H \rangle +| V \rangle)\xrightarrow{SLM2(\phi)}\frac{1}{\sqrt{2}}(|H\rangle+e^{i\phi}|V\rangle)\xrightarrow{HWP2}\frac{1}{2}((1+e^{i\phi})|H\rangle +(1-e^{i\phi})|V\rangle)\xrightarrow{PBS2}(1-e^{i\phi})|H\rangle $. Through the post-selection of PBS2, we can individually adjust the intensity of each beam. The phase generated by the beam intensity regulator is compensated by SLM1.}
\label{fig:schematic}
\end{figure*}

In this article we contribute the above challenges by creating an optical setup that is capable of (i) creating states close to the maximally entangled state, (ii) that are easy to interface with other means of distribution \cite{ding2017high} and (iii) can be fully explored via all two-dimensional subspace measurements. Of course, this setup comes with its own limitations and challenges, such as the manufacturing limit of beam displacers we employ and the number of two-dimensional subspaces in a large system. We develop adapted certification techniques that reveal that our setup achieves record fidelities in 32-dimensional entanglement ($F=0.933\pm0.001$), Schmidt numbers (30) and entangled bits (e-bits) as quantified by entanglement of formation ($E_{oF}=3.728\pm0.006$).

To start, let us describe our experimental setup, which is detailed in Fig.~\ref{fig:schematic}. As depicted in Fig.~\ref{fig:schematic}a, we divide the pumped light through beam displacers (BDs) into many parallel beams with the same energy (all half-wave plates (HWPs) in $404~nm$ are set to $22.5^{\circ}$.). Then every beam coherently induces spontaneous parametric down-conversion (SPDC) ($|H\rangle_{404nm}\longrightarrow|H\rangle_{808nm}\otimes|V\rangle_{808nm}$) on a BBO crystal. In particular, we use a collinear BBO crystal of type II. A polarising beam splitter (PBS) is then used to distribute the different polarisations along two different paths. Further beam displacers (BD1-BD5) are then used to form a $4\times8$ beam array. The spacing of each beam is $2~mm$ and the diameter of the pump beam is $0.6~mm$. Since, there are $32$ beams of light with equal pumping, our theoretical target state is $|\Phi^+\rangle=\frac{1}{\sqrt{32}} \sum_{i=0}^{31}|ii\rangle $, i.e. should be maximally entangled in all dimensions. 

\begin{figure*}[tbph]
\includegraphics [width= 1\textwidth]{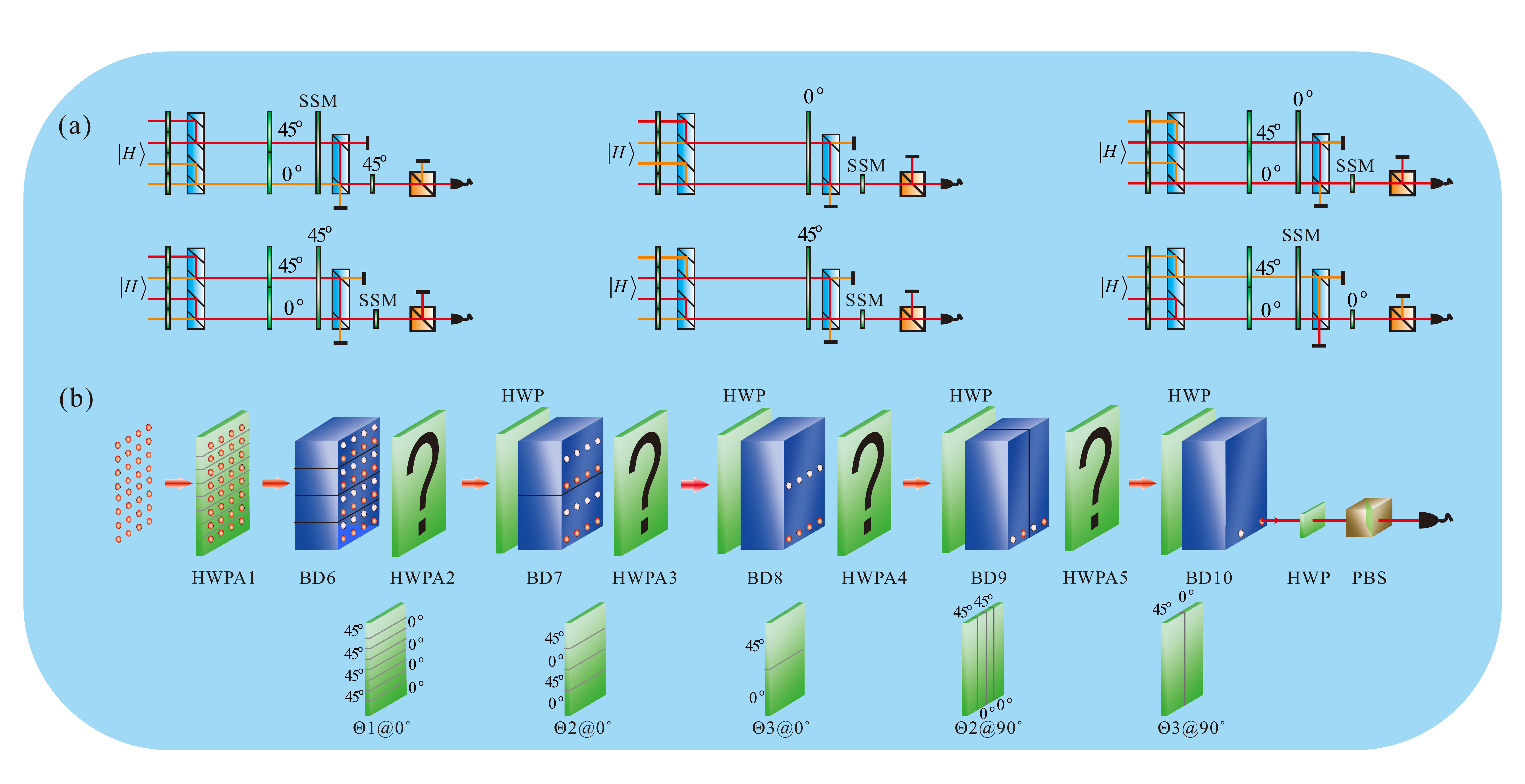}
\caption{Illustration of the two dimensional subspace measurement setup. (a) An exemplary $4$-dimensional measurement setup. There are $6$ necessary two-dimensional subspace measurements (SSM) of a $4$-dimensional path encoded state. By choosing the suitable combination of HWPAs, we can realise all of them. The red line represents the measured subspace. (b) An exemplary 32-dimensional subspace measurement setup. Similar to the $4$-dimensional setup, $\theta1$, $\theta2$, $\theta3$ and the HWP need to be chosen appropriately. Here, $\theta1 @ 0^{\circ}$, $\theta2 @ 0^{\circ}$ and $\theta3 @0^{\circ}$ represent HWPAs setting at $0^{\circ}$; $\theta 2 @90^{\circ}$ and $\theta 3@90^{\circ}$ represent HWPAs setting at $90^{\circ}$ and HWP represents a whole half-wave plate. }
\label{fig:measurement}
\end{figure*}

It is worth noting that our scheme is not simply an extension of Ref.~\cite{PhysRevLett.117.170403}, where also parallel beams were used to induce SPDC, as there are two distinct and significant advancements to report.
The first is that by using collinear BBO crystals, we can form two-dimensional beam arrays without the needed for spatial compensation. As this is one of the most demanding tasks, that increases in difficulty with increased dimension, we remove one of the central challenges to make our scheme scalable to even higher dimensions.

The second advancement concerns our BDs. Instead of using the more common crystals operating at $404~nm$ ($808~nm$), which in Ref.~\cite{PhysRevLett.117.170403} are approximately $36.41~mm$ ($39.70~mm$) long, thus introducing $4.21~mm$ separation between the horizontally and vertically polarised photons at $404~nm$ ($808~nm$), we manufactured new BDs from scratch. The main disadvantage of a natural crystal is its relatively strong absorption of ultraviolet light, and its relatively large volume. This was another challenge to high-dimensional scalability, which we overcame by designing and manufacturing a tailored structure, which is essentially a stacking of PBS and can achieve the same effect as a natural crystal BD (see also Fig.~\ref{fig:schematic}).

To compensate for the phase between the different paths used to encode the entangled state, we use a spatial light modulator (SLM), which can add an arbitrary phase to the vertically polarised light. We divide the SLM1 into $32$ pieces, each of them adjusts the phase of a beam of light. 
Because of imperfections in the manufactured BDs, the collection of light in each initial path can be different, which would lead to different amplitudes for the terms ($|ii\rangle$) of the high-dimensional entangled state. In order to compensate for this deviation, we also designed a beam intensity modulator. As shown in Fig.~\ref{fig:schematic}c, the setup consists of an SLM2, two HWP1-2 and a PBS2. The intensity of each beam can be adjusted by loading the phase on the second SLM. The counting rate of the entangled source is $\approx4000/s$, and the average coincidence efficiency is $0.16$. 

For the measurement setup, we use HWPAs and BDs to control each path using polarisation. We choose the path basis as computational basis, and thus measurements of the diagonal elements of the density matrix $\langle ij|\rho|ij\rangle$ are straightforward (using a coincidence logic with detectors in the paths). To certify entanglement or the actual fidelity with our target state, we additionally need access to the real parts of the following off-diagonal elements $\Re e[\langle i i|\rho| j j\rangle]$. This requires two mutually unbiased measurements in all two dimensional subspaces, labeled by $i$ and $j$.  As $\Re e[\langle ii|\rho|jj\rangle]=\Re e[\langle jj|\rho|ii\rangle]$, we need $496$ unique combinations to measure all real parts with $i,j=0,1,\dots,31 (i<j)$. 

In our measurement setup, any two-dimensional subspace measurement can easily be realised. In Fig.~\ref{fig:measurement}a, we show the state measurement setup in an exemplary $4$-dimensional case. By setting up each HWPA appropriately, we can realise the measurement in any 2-dimensional subspaces of the $4\times4$ dimensional system. This is easy to generalise and scale, which we showcase in Fig.~\ref{fig:measurement}b, where we only need to set HWPA2-HWPA5 appropriately, and then we can measure in all $2$-dimensional subspaces of our 32-dimensional system. Fortunately, only a few HWPAs are needed to measure arbitrary two-dimensional subspaces. We showcase part of the HWPA settings in Appendix E.

\begin{figure*}[tbph]
\centering
\includegraphics [width= 1\textwidth]{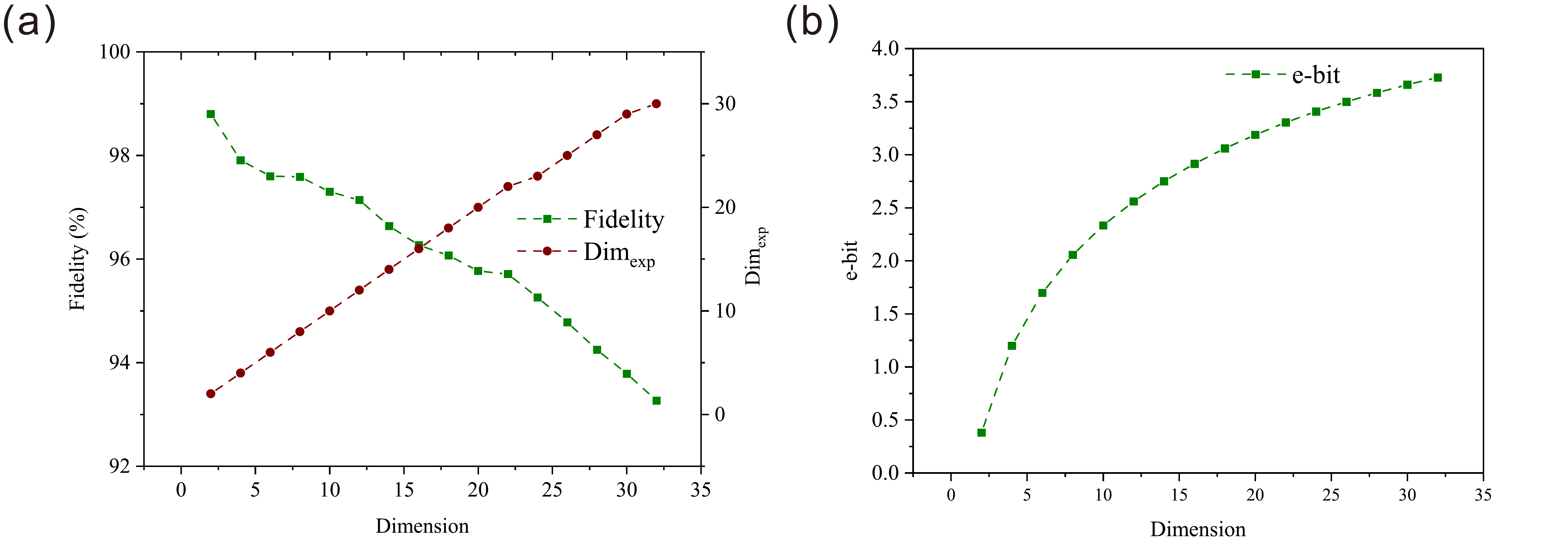}
\caption{Experimental results for fidelity $F_+$, resulting Schmidt number $\text{Dim}_{\text{exp}}$ and e-bit ($E_{oF}$). (a) The results of fidelity and Schmidt number. The green line represents the fidelity as a function of the physical dimension, and the red line represents the entangled dimension (Schmidt number) witnessed according to the fidelity. We achieve a fidelity of $F_{d=32}=(93.3\pm0.1)\%>29/32$ with a $32$ dimensional maximally entangled state, which proves our state features at least Schmidt number $30$ in $32$ dimensions. (b) The resulting $E_{oF}$. The green line represents the entanglement of formation values for each dimension. When $d=32$, it reaches a maximum value of $B_{d=22}=3.728\pm0.006$. Error bars are in the range of points and thus not visible.}
\label{fig:result}
\end{figure*}

For the target state fidelity ${F_+}:=\text{Tr}(\rho_{exp}|\Phi^+\rangle\langle\Phi^+|)$ it is actually sufficient to measure $32$ out of the $1024$ diagonal terms, i.e. just the terms $\langle ii|\rho_{exp}|ii\rangle$. We thus focus on their measurements and develop entanglement quantification methods that can be used without access to the terms $\langle ij|\rho|ij\rangle(i\neq j)$. In any case, because the isolation of each linear device is more than $10^{-3}$, and we have measured $\langle 0j|\rho|0j\rangle$ and confirmed that the maximal value is $4.49\times10^{-5}$, we can estimate $\langle ij|\rho|ij\rangle^{i\neq j}\approx4.49\times10^{-5}$. 

\emph{Experimental Results} We now proceed to present the experimental results of the setup. We performed $4000$-single outcome projective measurements using two detectors and a coincidence logic at a coincidence window of $3~ns$, significantly less than the $~10^6$ such measurements needed for a full state tomography. This allowed us to accurately measure all elements $\{\langle ii|\rho|jj\rangle\}$, in which $i,j\in\{0,1,\cdots,31\}$. We should also say that we take the raw coincidences without any further background subtraction, which would greatly improve our values, but run against the spirit of certification. This yields the following fidelity:
\begin{align}
F_{+}=\frac{1}{32}\sum_{i,j=0}^{31}\langle ii|\rho_{exp}| jj \rangle.
\end{align}
As we can see, the decrease of fidelity seems to be almost linear in this regime. The white noise resistance for the maximally entangled state  $p|\Phi^+\rangle\langle\Phi^+|+(1-p)\frac{\mathbbm{1}}{d^2}$, goes as ${F}_{sep}=\frac{1}{d}$, i.e. the point of separability is reached with a scaling of $1/d$, so in our experiment the gap between reached fidelity and separability still grows larger with each dimension. This is very promising for scaling up and getting an advantage for even higher dimensions in our scheme. Furthermore, the fidelity allows for a direct lower bound on the Schmidt number, as the critical fidelity for Schmidt number $k$ is ${F}_{k}=\frac{k}{d}$, which proves our state features at least Schmidt number $30$ in $32$ dimensions with 32 standard deviations confidence. These are to date the highest achieved fidelities for high-dimensional states, together with the very recent Ref.\cite{valencia2020highdimensional}, which produces similar fidelities in a pixel basis, corroborating the suitability of the spatial domain for high-dimensional entanglement.
To go beyond fidelity as an entanglement characterisation (which comes with its own limitations \cite{weilenmann2019entanglement}), we also develop a method to estimate entanglement of formation \cite{Hill_1997}, from our data. Roughly speaking, it provides a means of quantifying how much entanglement is needed to create the state and can be defined as the average entanglement entropy minimised over all decompositions $E_{oF}:=\inf_{\mathcal{D}}(\rho)\sum_ip_iS(\text{Tr}_B(|\psi_i\rangle\langle\psi_i|)$. Just as in \cite{Schneeloch_2018,Schneeloch_2019}, we use the conditional quantum uncertainty relation from \cite{Berta_2010} to estimate negative conditional entropy, which is a lower bound to entanglement of formation \cite{Carlen_2012}. In its usual form, however, one would need access to the full statistics of two mutually unbiased measurements, which we did not perform. Using some additional bounds and estimates, it is, however, possible to provide a lower bound on the negative conditional entropy from diagonal correlations and two-dimensional subspace measurements. The bounds rely on the invariance of $|\Phi^+\rangle$ under local unitary rotations $U\otimes U^*$, which can be used to lower bound joint entropies in the mutually unbiased basis without having to measure them directly. We provide further details of the method in  appendix B. Using this method we can certify a lower bound on the entanglement of formation of $E_{oF}\geq 3.728\pm0.006$. All results are shown in Fig.~\ref{fig:result}. As a comparison, the record number in time-bin implementations is $2.09$ in dimension $7$ \cite{Martin_2017} and in spatial modes $3.43$ in $262144$ dimensions \cite{Schneeloch_2019} (this is comparing to equal paradigms, i.e. without additional assumptions or background subtraction).

\emph{Conclusion} We have presented a framework for high-dimensional photonic entanglement to be generated in spatial modes. Moreover, we present a scalable measurement architecture that is capable of reading out coherences in any two-dimensional subspace. Our experimental implementation not only shows the feasibility of the approach, but also displays the highest quality entanglement in high-dimensions created in optical setups thus far, making it a prime candidate for realising the promising advantages of high-dimensional entanglement in quantum communication. To analyse the state, we have used single-outcome measurements, i.e. one detector per side with coincidence post-selection. Contrary to OAM entanglement, where this is a feature of post-selecting with a single-mode fibre, there is no fundamental reason for this in our setup. By including more detectors, one could, without any mode-dependent loss, include many more outcomes per side, enabling device independent tests in higher dimensions and truly improving key rates in high-dimensional key distribution. For that purpose, in appendix D, we also include a theoretical setup that would allow for full scale mutually unbiased measurements in our multi-path entangled source. The array design of our experimental device lets all the spatial modes keep a high degree of similarity, which is conducive to the combination of multi-core fibre to complete the long-distance distribution of high-dimensional entanglement, forming a high-dimensional quantum network \cite{ding2017high}. As the number of two-dimensional subspaces grows quadratically in the dimension, it may be useful to explore other measurement regimes, such as mutually unbiased bases or complementing some subspaces by semi-definite programming techniques to estimate unmeasured subspaces as in \cite{Martin_2017}. 

\emph{Acknowledgements} This work was supported by the National Key Research and Development Program of China (No.\ 2017YFA0304100, No. 2016YFA0301300 and No. 2016YFA0301700), NSFC (Nos. 11774335, 11734015, 11874345, 11821404, 11904357), the Key Research Program of Frontier Sciences, CAS (No.\ QYZDY-SSW-SLH003), Science Foundation of the CAS (ZDRW-XH-2019-1), the Fundamental Research Funds for the Central Universities, and Anhui Initiative in Quantum Information Technologies (Nos.\ AHY020100, AHY060300). MH and PE acknowledge funding from the Austrian Science Fund (FWF) through the STARTproject Y879-N27 and the Zukunftskolleg ZK03.
\bibliography{bibliography}

\begin{thebibliography}{30}
\expandafter\ifx\csname natexlab\endcsname\relax\def\natexlab#1{#1}\fi
\expandafter\ifx\csname bibnamefont\endcsname\relax
  \def\bibnamefont#1{#1}\fi
\expandafter\ifx\csname bibfnamefont\endcsname\relax
  \def\bibfnamefont#1{#1}\fi
\expandafter\ifx\csname citenamefont\endcsname\relax
  \def\citenamefont#1{#1}\fi
\expandafter\ifx\csname url\endcsname\relax
  \def\url#1{\texttt{#1}}\fi
\expandafter\ifx\csname urlprefix\endcsname\relax\def\urlprefix{URL }\fi
\providecommand{\bibinfo}[2]{#2}
\providecommand{\eprint}[2][]{\url{#2}}

\bibitem[{\citenamefont{Friis et~al.}(2018)\citenamefont{Friis, Vitagliano,
  Malik, and Huber}}]{Friis_2018}
\bibinfo{author}{\bibfnamefont{N.}~\bibnamefont{Friis}},
  \bibinfo{author}{\bibfnamefont{G.}~\bibnamefont{Vitagliano}},
  \bibinfo{author}{\bibfnamefont{M.}~\bibnamefont{Malik}}, \bibnamefont{and}
  \bibinfo{author}{\bibfnamefont{M.}~\bibnamefont{Huber}},
  \bibinfo{journal}{Nature Reviews Physics} \textbf{\bibinfo{volume}{1}},
  \bibinfo{pages}{72–87} (\bibinfo{year}{2018}), ISSN
  \bibinfo{issn}{2522-5820},
  \urlprefix\url{http://dx.doi.org/10.1038/s42254-018-0003-5}.

\bibitem[{\citenamefont{Cozzolino
  et~al.}(2019{\natexlab{a}})\citenamefont{Cozzolino, Da~Lio, Bacco, and
  Oxenløwe}}]{Cozzolino_2019b}
\bibinfo{author}{\bibfnamefont{D.}~\bibnamefont{Cozzolino}},
  \bibinfo{author}{\bibfnamefont{B.}~\bibnamefont{Da~Lio}},
  \bibinfo{author}{\bibfnamefont{D.}~\bibnamefont{Bacco}}, \bibnamefont{and}
  \bibinfo{author}{\bibfnamefont{L.~K.} \bibnamefont{Oxenløwe}},
  \bibinfo{journal}{Advanced Quantum Technologies}
  \textbf{\bibinfo{volume}{2}}, \bibinfo{pages}{1900038}
  (\bibinfo{year}{2019}{\natexlab{a}}), ISSN \bibinfo{issn}{2511-9044},
  \urlprefix\url{http://dx.doi.org/10.1002/qute.201900038}.

\bibitem[{\citenamefont{Tiranov et~al.}(2017)\citenamefont{Tiranov, Designolle,
  Cruzeiro, Lavoie, Brunner, Afzelius, Huber, and Gisin}}]{Tiranov_2017}
\bibinfo{author}{\bibfnamefont{A.}~\bibnamefont{Tiranov}},
  \bibinfo{author}{\bibfnamefont{S.}~\bibnamefont{Designolle}},
  \bibinfo{author}{\bibfnamefont{E.~Z.} \bibnamefont{Cruzeiro}},
  \bibinfo{author}{\bibfnamefont{J.}~\bibnamefont{Lavoie}},
  \bibinfo{author}{\bibfnamefont{N.}~\bibnamefont{Brunner}},
  \bibinfo{author}{\bibfnamefont{M.}~\bibnamefont{Afzelius}},
  \bibinfo{author}{\bibfnamefont{M.}~\bibnamefont{Huber}}, \bibnamefont{and}
  \bibinfo{author}{\bibfnamefont{N.}~\bibnamefont{Gisin}},
  \bibinfo{journal}{Physical Review A} \textbf{\bibinfo{volume}{96}}
  (\bibinfo{year}{2017}), ISSN \bibinfo{issn}{2469-9934},
  \urlprefix\url{http://dx.doi.org/10.1103/PhysRevA.96.040303}.

\bibitem[{\citenamefont{Martin et~al.}(2017)\citenamefont{Martin, Guerreiro,
  Tiranov, Designolle, Fröwis, Brunner, Huber, and Gisin}}]{Martin_2017}
\bibinfo{author}{\bibfnamefont{A.}~\bibnamefont{Martin}},
  \bibinfo{author}{\bibfnamefont{T.}~\bibnamefont{Guerreiro}},
  \bibinfo{author}{\bibfnamefont{A.}~\bibnamefont{Tiranov}},
  \bibinfo{author}{\bibfnamefont{S.}~\bibnamefont{Designolle}},
  \bibinfo{author}{\bibfnamefont{F.}~\bibnamefont{Fröwis}},
  \bibinfo{author}{\bibfnamefont{N.}~\bibnamefont{Brunner}},
  \bibinfo{author}{\bibfnamefont{M.}~\bibnamefont{Huber}}, \bibnamefont{and}
  \bibinfo{author}{\bibfnamefont{N.}~\bibnamefont{Gisin}},
  \bibinfo{journal}{Physical Review Letters} \textbf{\bibinfo{volume}{118}}
  (\bibinfo{year}{2017}), ISSN \bibinfo{issn}{1079-7114},
  \urlprefix\url{http://dx.doi.org/10.1103/PhysRevLett.118.110501}.

\bibitem[{\citenamefont{Dada et~al.}(2011)\citenamefont{Dada, Leach, Buller,
  Padgett, and Andersson}}]{Dada:2011vc}
\bibinfo{author}{\bibfnamefont{A.~C.} \bibnamefont{Dada}},
  \bibinfo{author}{\bibfnamefont{J.}~\bibnamefont{Leach}},
  \bibinfo{author}{\bibfnamefont{G.~S.} \bibnamefont{Buller}},
  \bibinfo{author}{\bibfnamefont{M.~J.} \bibnamefont{Padgett}},
  \bibnamefont{and}
  \bibinfo{author}{\bibfnamefont{E.}~\bibnamefont{Andersson}},
  \bibinfo{journal}{Nature Physics} \textbf{\bibinfo{volume}{7}},
  \bibinfo{pages}{677} (\bibinfo{year}{2011}).

\bibitem[{\citenamefont{Krenn et~al.}(2017)\citenamefont{Krenn, Malik, Erhard,
  and Zeilinger}}]{Krenn:2017hz}
\bibinfo{author}{\bibfnamefont{M.}~\bibnamefont{Krenn}},
  \bibinfo{author}{\bibfnamefont{M.}~\bibnamefont{Malik}},
  \bibinfo{author}{\bibfnamefont{M.}~\bibnamefont{Erhard}}, \bibnamefont{and}
  \bibinfo{author}{\bibfnamefont{A.}~\bibnamefont{Zeilinger}},
  \bibinfo{journal}{Phil. Trans. R. Soc. A} \textbf{\bibinfo{volume}{375}},
  \bibinfo{pages}{20150442} (\bibinfo{year}{2017}).

\bibitem[{\citenamefont{Bavaresco et~al.}(2018)\citenamefont{Bavaresco,
  Herrera~Valencia, Klöckl, Pivoluska, Erker, Friis, Malik, and
  Huber}}]{Bavaresco_2018}
\bibinfo{author}{\bibfnamefont{J.}~\bibnamefont{Bavaresco}},
  \bibinfo{author}{\bibfnamefont{N.}~\bibnamefont{Herrera~Valencia}},
  \bibinfo{author}{\bibfnamefont{C.}~\bibnamefont{Klöckl}},
  \bibinfo{author}{\bibfnamefont{M.}~\bibnamefont{Pivoluska}},
  \bibinfo{author}{\bibfnamefont{P.}~\bibnamefont{Erker}},
  \bibinfo{author}{\bibfnamefont{N.}~\bibnamefont{Friis}},
  \bibinfo{author}{\bibfnamefont{M.}~\bibnamefont{Malik}}, \bibnamefont{and}
  \bibinfo{author}{\bibfnamefont{M.}~\bibnamefont{Huber}},
  \bibinfo{journal}{Nature Physics} \textbf{\bibinfo{volume}{14}},
  \bibinfo{pages}{1032–1037} (\bibinfo{year}{2018}), ISSN
  \bibinfo{issn}{1745-2481},
  \urlprefix\url{http://dx.doi.org/10.1038/s41567-018-0203-z}.

\bibitem[{\citenamefont{Kues et~al.}(2017)\citenamefont{Kues, Reimer, Roztocki,
  Cort{\'e}s, Sciara, Wetzel, Zhang, Cino, Chu, Little et~al.}}]{kues2017chip}
\bibinfo{author}{\bibfnamefont{M.}~\bibnamefont{Kues}},
  \bibinfo{author}{\bibfnamefont{C.}~\bibnamefont{Reimer}},
  \bibinfo{author}{\bibfnamefont{P.}~\bibnamefont{Roztocki}},
  \bibinfo{author}{\bibfnamefont{L.~R.} \bibnamefont{Cort{\'e}s}},
  \bibinfo{author}{\bibfnamefont{S.}~\bibnamefont{Sciara}},
  \bibinfo{author}{\bibfnamefont{B.}~\bibnamefont{Wetzel}},
  \bibinfo{author}{\bibfnamefont{Y.}~\bibnamefont{Zhang}},
  \bibinfo{author}{\bibfnamefont{A.}~\bibnamefont{Cino}},
  \bibinfo{author}{\bibfnamefont{S.~T.} \bibnamefont{Chu}},
  \bibinfo{author}{\bibfnamefont{B.~E.} \bibnamefont{Little}},
  \bibnamefont{et~al.}, \bibinfo{journal}{Nature}
  \textbf{\bibinfo{volume}{546}}, \bibinfo{pages}{622} (\bibinfo{year}{2017}).

\bibitem[{\citenamefont{Wang et~al.}(2018)\citenamefont{Wang, Paesani, Ding,
  Santagati, Skrzypczyk, Salavrakos, Tura, Augusiak, Mančinska, Bacco
  et~al.}}]{Wang_2018}
\bibinfo{author}{\bibfnamefont{J.}~\bibnamefont{Wang}},
  \bibinfo{author}{\bibfnamefont{S.}~\bibnamefont{Paesani}},
  \bibinfo{author}{\bibfnamefont{Y.}~\bibnamefont{Ding}},
  \bibinfo{author}{\bibfnamefont{R.}~\bibnamefont{Santagati}},
  \bibinfo{author}{\bibfnamefont{P.}~\bibnamefont{Skrzypczyk}},
  \bibinfo{author}{\bibfnamefont{A.}~\bibnamefont{Salavrakos}},
  \bibinfo{author}{\bibfnamefont{J.}~\bibnamefont{Tura}},
  \bibinfo{author}{\bibfnamefont{R.}~\bibnamefont{Augusiak}},
  \bibinfo{author}{\bibfnamefont{L.}~\bibnamefont{Mančinska}},
  \bibinfo{author}{\bibfnamefont{D.}~\bibnamefont{Bacco}},
  \bibnamefont{et~al.}, \bibinfo{journal}{Science}
  \textbf{\bibinfo{volume}{360}}, \bibinfo{pages}{285–291}
  (\bibinfo{year}{2018}), ISSN \bibinfo{issn}{1095-9203},
  \urlprefix\url{http://dx.doi.org/10.1126/science.aar7053}.

\bibitem[{\citenamefont{Llewellyn et~al.}(2019)\citenamefont{Llewellyn, Ding,
  Faruque, Paesani, Bacco, Santagati, Qian, Li, Xiao, Huber
  et~al.}}]{Llewellyn_2019}
\bibinfo{author}{\bibfnamefont{D.}~\bibnamefont{Llewellyn}},
  \bibinfo{author}{\bibfnamefont{Y.}~\bibnamefont{Ding}},
  \bibinfo{author}{\bibfnamefont{I.~I.} \bibnamefont{Faruque}},
  \bibinfo{author}{\bibfnamefont{S.}~\bibnamefont{Paesani}},
  \bibinfo{author}{\bibfnamefont{D.}~\bibnamefont{Bacco}},
  \bibinfo{author}{\bibfnamefont{R.}~\bibnamefont{Santagati}},
  \bibinfo{author}{\bibfnamefont{Y.-J.} \bibnamefont{Qian}},
  \bibinfo{author}{\bibfnamefont{Y.}~\bibnamefont{Li}},
  \bibinfo{author}{\bibfnamefont{Y.-F.} \bibnamefont{Xiao}},
  \bibinfo{author}{\bibfnamefont{M.}~\bibnamefont{Huber}},
  \bibnamefont{et~al.}, \bibinfo{journal}{Nature Physics}
  \textbf{\bibinfo{volume}{16}}, \bibinfo{pages}{148–153}
  (\bibinfo{year}{2019}), ISSN \bibinfo{issn}{1745-2481},
  \urlprefix\url{http://dx.doi.org/10.1038/s41567-019-0727-x}.

\bibitem[{\citenamefont{Cozzolino
  et~al.}(2019{\natexlab{b}})\citenamefont{Cozzolino, Bacco, Da~Lio, Ingerslev,
  Ding, Dalgaard, Kristensen, Galili, Rottwitt, Ramachandran
  et~al.}}]{Cozzolino_2019}
\bibinfo{author}{\bibfnamefont{D.}~\bibnamefont{Cozzolino}},
  \bibinfo{author}{\bibfnamefont{D.}~\bibnamefont{Bacco}},
  \bibinfo{author}{\bibfnamefont{B.}~\bibnamefont{Da~Lio}},
  \bibinfo{author}{\bibfnamefont{K.}~\bibnamefont{Ingerslev}},
  \bibinfo{author}{\bibfnamefont{Y.}~\bibnamefont{Ding}},
  \bibinfo{author}{\bibfnamefont{K.}~\bibnamefont{Dalgaard}},
  \bibinfo{author}{\bibfnamefont{P.}~\bibnamefont{Kristensen}},
  \bibinfo{author}{\bibfnamefont{M.}~\bibnamefont{Galili}},
  \bibinfo{author}{\bibfnamefont{K.}~\bibnamefont{Rottwitt}},
  \bibinfo{author}{\bibfnamefont{S.}~\bibnamefont{Ramachandran}},
  \bibnamefont{et~al.}, \bibinfo{journal}{Physical Review Applied}
  \textbf{\bibinfo{volume}{11}} (\bibinfo{year}{2019}{\natexlab{b}}), ISSN
  \bibinfo{issn}{2331-7019},
  \urlprefix\url{http://dx.doi.org/10.1103/PhysRevApplied.11.064058}.

\bibitem[{\citenamefont{Cao et~al.}(2020)\citenamefont{Cao, Gao, Zhang, Wang,
  He, Liu, Zhou, Chen, Li, Yu et~al.}}]{cao2020distribution}
\bibinfo{author}{\bibfnamefont{H.}~\bibnamefont{Cao}},
  \bibinfo{author}{\bibfnamefont{S.-C.} \bibnamefont{Gao}},
  \bibinfo{author}{\bibfnamefont{C.}~\bibnamefont{Zhang}},
  \bibinfo{author}{\bibfnamefont{J.}~\bibnamefont{Wang}},
  \bibinfo{author}{\bibfnamefont{D.-Y.} \bibnamefont{He}},
  \bibinfo{author}{\bibfnamefont{B.-H.} \bibnamefont{Liu}},
  \bibinfo{author}{\bibfnamefont{Z.-W.} \bibnamefont{Zhou}},
  \bibinfo{author}{\bibfnamefont{Y.-J.} \bibnamefont{Chen}},
  \bibinfo{author}{\bibfnamefont{Z.-H.} \bibnamefont{Li}},
  \bibinfo{author}{\bibfnamefont{S.-Y.} \bibnamefont{Yu}},
  \bibnamefont{et~al.}, \bibinfo{journal}{Optica} \textbf{\bibinfo{volume}{7}},
  \bibinfo{pages}{232} (\bibinfo{year}{2020}).

\bibitem[{\citenamefont{Valencia et~al.}(2019)\citenamefont{Valencia, Goel,
  McCutcheon, Defienne, and Malik}}]{Valencia:2019tc}
\bibinfo{author}{\bibfnamefont{N.~H.} \bibnamefont{Valencia}},
  \bibinfo{author}{\bibfnamefont{S.}~\bibnamefont{Goel}},
  \bibinfo{author}{\bibfnamefont{W.}~\bibnamefont{McCutcheon}},
  \bibinfo{author}{\bibfnamefont{H.}~\bibnamefont{Defienne}}, \bibnamefont{and}
  \bibinfo{author}{\bibfnamefont{M.}~\bibnamefont{Malik}},
  \bibinfo{journal}{arXiv} p. \bibinfo{pages}{1910.04490}
  (\bibinfo{year}{2019}), \eprint{1910.04490}.

\bibitem[{\citenamefont{Schaeff et~al.}(2015)\citenamefont{Schaeff, Polster,
  Huber, Ramelow, and Zeilinger}}]{Schaeff_2015}
\bibinfo{author}{\bibfnamefont{C.}~\bibnamefont{Schaeff}},
  \bibinfo{author}{\bibfnamefont{R.}~\bibnamefont{Polster}},
  \bibinfo{author}{\bibfnamefont{M.}~\bibnamefont{Huber}},
  \bibinfo{author}{\bibfnamefont{S.}~\bibnamefont{Ramelow}}, \bibnamefont{and}
  \bibinfo{author}{\bibfnamefont{A.}~\bibnamefont{Zeilinger}},
  \bibinfo{journal}{Optica} \textbf{\bibinfo{volume}{2}}, \bibinfo{pages}{523}
  (\bibinfo{year}{2015}), ISSN \bibinfo{issn}{2334-2536},
  \urlprefix\url{http://dx.doi.org/10.1364/OPTICA.2.000523}.

\bibitem[{\citenamefont{Bouchard et~al.}(2018)\citenamefont{Bouchard, Valencia,
  Brandt, Fickler, Huber, and Malik}}]{Bouchard_2018}
\bibinfo{author}{\bibfnamefont{F.}~\bibnamefont{Bouchard}},
  \bibinfo{author}{\bibfnamefont{N.~H.} \bibnamefont{Valencia}},
  \bibinfo{author}{\bibfnamefont{F.}~\bibnamefont{Brandt}},
  \bibinfo{author}{\bibfnamefont{R.}~\bibnamefont{Fickler}},
  \bibinfo{author}{\bibfnamefont{M.}~\bibnamefont{Huber}}, \bibnamefont{and}
  \bibinfo{author}{\bibfnamefont{M.}~\bibnamefont{Malik}},
  \bibinfo{journal}{Optics Express} \textbf{\bibinfo{volume}{26}},
  \bibinfo{pages}{31925} (\bibinfo{year}{2018}), ISSN
  \bibinfo{issn}{1094-4087},
  \urlprefix\url{http://dx.doi.org/10.1364/OE.26.031925}.

\bibitem[{\citenamefont{Brendel et~al.}(1999)\citenamefont{Brendel, Gisin,
  Tittel, and Zbinden}}]{PhysRevLett.82.2594}
\bibinfo{author}{\bibfnamefont{J.}~\bibnamefont{Brendel}},
  \bibinfo{author}{\bibfnamefont{N.}~\bibnamefont{Gisin}},
  \bibinfo{author}{\bibfnamefont{W.}~\bibnamefont{Tittel}}, \bibnamefont{and}
  \bibinfo{author}{\bibfnamefont{H.}~\bibnamefont{Zbinden}},
  \bibinfo{journal}{Phys. Rev. Lett.} \textbf{\bibinfo{volume}{82}},
  \bibinfo{pages}{2594} (\bibinfo{year}{1999}),
  \urlprefix\url{https://link.aps.org/doi/10.1103/PhysRevLett.82.2594}.

\bibitem[{\citenamefont{Olislager et~al.}(2012)\citenamefont{Olislager, Mbodji,
  Woodhead, Cussey, Furfaro, Emplit, Massar, Huy, and
  Merolla}}]{Olislager_2012}
\bibinfo{author}{\bibfnamefont{L.}~\bibnamefont{Olislager}},
  \bibinfo{author}{\bibfnamefont{I.}~\bibnamefont{Mbodji}},
  \bibinfo{author}{\bibfnamefont{E.}~\bibnamefont{Woodhead}},
  \bibinfo{author}{\bibfnamefont{J.}~\bibnamefont{Cussey}},
  \bibinfo{author}{\bibfnamefont{L.}~\bibnamefont{Furfaro}},
  \bibinfo{author}{\bibfnamefont{P.}~\bibnamefont{Emplit}},
  \bibinfo{author}{\bibfnamefont{S.}~\bibnamefont{Massar}},
  \bibinfo{author}{\bibfnamefont{K.~P.} \bibnamefont{Huy}}, \bibnamefont{and}
  \bibinfo{author}{\bibfnamefont{J.-M.} \bibnamefont{Merolla}},
  \bibinfo{journal}{New Journal of Physics} \textbf{\bibinfo{volume}{14}},
  \bibinfo{pages}{043015} (\bibinfo{year}{2012}),
  \urlprefix\url{https://doi.org/10.1088%2F1367-2630%2F14%2F4%2F043015}.

\bibitem[{\citenamefont{Chen et~al.}(2020)\citenamefont{Chen, Ecker, Bavaresco,
  Scheidl, Chen, Steinlechner, Huber, and Ursin}}]{Chen_2020}
\bibinfo{author}{\bibfnamefont{Y.}~\bibnamefont{Chen}},
  \bibinfo{author}{\bibfnamefont{S.}~\bibnamefont{Ecker}},
  \bibinfo{author}{\bibfnamefont{J.}~\bibnamefont{Bavaresco}},
  \bibinfo{author}{\bibfnamefont{T.}~\bibnamefont{Scheidl}},
  \bibinfo{author}{\bibfnamefont{L.}~\bibnamefont{Chen}},
  \bibinfo{author}{\bibfnamefont{F.}~\bibnamefont{Steinlechner}},
  \bibinfo{author}{\bibfnamefont{M.}~\bibnamefont{Huber}}, \bibnamefont{and}
  \bibinfo{author}{\bibfnamefont{R.}~\bibnamefont{Ursin}},
  \bibinfo{journal}{Physical Review A} \textbf{\bibinfo{volume}{101}}
  (\bibinfo{year}{2020}), ISSN \bibinfo{issn}{2469-9934},
  \urlprefix\url{http://dx.doi.org/10.1103/PhysRevA.101.032302}.

\bibitem[{\citenamefont{Steinlechner et~al.}(2017)\citenamefont{Steinlechner,
  Ecker, Fink, Liu, Bavaresco, Huber, Scheidl, and Ursin}}]{Steinlechner_2017}
\bibinfo{author}{\bibfnamefont{F.}~\bibnamefont{Steinlechner}},
  \bibinfo{author}{\bibfnamefont{S.}~\bibnamefont{Ecker}},
  \bibinfo{author}{\bibfnamefont{M.}~\bibnamefont{Fink}},
  \bibinfo{author}{\bibfnamefont{B.}~\bibnamefont{Liu}},
  \bibinfo{author}{\bibfnamefont{J.}~\bibnamefont{Bavaresco}},
  \bibinfo{author}{\bibfnamefont{M.}~\bibnamefont{Huber}},
  \bibinfo{author}{\bibfnamefont{T.}~\bibnamefont{Scheidl}}, \bibnamefont{and}
  \bibinfo{author}{\bibfnamefont{R.}~\bibnamefont{Ursin}},
  \bibinfo{journal}{Nature Communications} \textbf{\bibinfo{volume}{8}}
  (\bibinfo{year}{2017}), ISSN \bibinfo{issn}{2041-1723},
  \urlprefix\url{http://dx.doi.org/10.1038/ncomms15971}.

\bibitem[{\citenamefont{Ecker et~al.}(2019)\citenamefont{Ecker, Bouchard,
  Bulla, Brandt, Kohout, Steinlechner, Fickler, Malik, Guryanova, Ursin
  et~al.}}]{Ecker_2019}
\bibinfo{author}{\bibfnamefont{S.}~\bibnamefont{Ecker}},
  \bibinfo{author}{\bibfnamefont{F.}~\bibnamefont{Bouchard}},
  \bibinfo{author}{\bibfnamefont{L.}~\bibnamefont{Bulla}},
  \bibinfo{author}{\bibfnamefont{F.}~\bibnamefont{Brandt}},
  \bibinfo{author}{\bibfnamefont{O.}~\bibnamefont{Kohout}},
  \bibinfo{author}{\bibfnamefont{F.}~\bibnamefont{Steinlechner}},
  \bibinfo{author}{\bibfnamefont{R.}~\bibnamefont{Fickler}},
  \bibinfo{author}{\bibfnamefont{M.}~\bibnamefont{Malik}},
  \bibinfo{author}{\bibfnamefont{Y.}~\bibnamefont{Guryanova}},
  \bibinfo{author}{\bibfnamefont{R.}~\bibnamefont{Ursin}},
  \bibnamefont{et~al.}, \bibinfo{journal}{Physical Review X}
  \textbf{\bibinfo{volume}{9}} (\bibinfo{year}{2019}), ISSN
  \bibinfo{issn}{2160-3308},
  \urlprefix\url{http://dx.doi.org/10.1103/PhysRevX.9.041042}.

\bibitem[{\citenamefont{Ding et~al.}(2017)\citenamefont{Ding, Bacco, Dalgaard,
  Cai, Zhou, Rottwitt, and Oxenl{\o}we}}]{ding2017high}
\bibinfo{author}{\bibfnamefont{Y.}~\bibnamefont{Ding}},
  \bibinfo{author}{\bibfnamefont{D.}~\bibnamefont{Bacco}},
  \bibinfo{author}{\bibfnamefont{K.}~\bibnamefont{Dalgaard}},
  \bibinfo{author}{\bibfnamefont{X.}~\bibnamefont{Cai}},
  \bibinfo{author}{\bibfnamefont{X.}~\bibnamefont{Zhou}},
  \bibinfo{author}{\bibfnamefont{K.}~\bibnamefont{Rottwitt}}, \bibnamefont{and}
  \bibinfo{author}{\bibfnamefont{L.~K.} \bibnamefont{Oxenl{\o}we}},
  \bibinfo{journal}{npj Quantum Information} \textbf{\bibinfo{volume}{3}},
  \bibinfo{pages}{1} (\bibinfo{year}{2017}).

\bibitem[{\citenamefont{Hu et~al.}(2016)\citenamefont{Hu, Chen, Liu, Guo,
  Huang, Zhou, Han, Li, and Guo}}]{PhysRevLett.117.170403}
\bibinfo{author}{\bibfnamefont{X.-M.} \bibnamefont{Hu}},
  \bibinfo{author}{\bibfnamefont{J.-S.} \bibnamefont{Chen}},
  \bibinfo{author}{\bibfnamefont{B.-H.} \bibnamefont{Liu}},
  \bibinfo{author}{\bibfnamefont{Y.}~\bibnamefont{Guo}},
  \bibinfo{author}{\bibfnamefont{Y.-F.} \bibnamefont{Huang}},
  \bibinfo{author}{\bibfnamefont{Z.-Q.} \bibnamefont{Zhou}},
  \bibinfo{author}{\bibfnamefont{Y.-J.} \bibnamefont{Han}},
  \bibinfo{author}{\bibfnamefont{C.-F.} \bibnamefont{Li}}, \bibnamefont{and}
  \bibinfo{author}{\bibfnamefont{G.-C.} \bibnamefont{Guo}},
  \bibinfo{journal}{Phys. Rev. Lett.} \textbf{\bibinfo{volume}{117}},
  \bibinfo{pages}{170403} (\bibinfo{year}{2016}),
  \urlprefix\url{https://link.aps.org/doi/10.1103/PhysRevLett.117.170403}.

\bibitem[{\citenamefont{Valencia et~al.}(2020)\citenamefont{Valencia,
  Srivastav, Pivoluska, Huber, Friis, McCutcheon, and
  Malik}}]{valencia2020highdimensional}
\bibinfo{author}{\bibfnamefont{N.~H.} \bibnamefont{Valencia}},
  \bibinfo{author}{\bibfnamefont{V.}~\bibnamefont{Srivastav}},
  \bibinfo{author}{\bibfnamefont{M.}~\bibnamefont{Pivoluska}},
  \bibinfo{author}{\bibfnamefont{M.}~\bibnamefont{Huber}},
  \bibinfo{author}{\bibfnamefont{N.}~\bibnamefont{Friis}},
  \bibinfo{author}{\bibfnamefont{W.}~\bibnamefont{McCutcheon}},
  \bibnamefont{and} \bibinfo{author}{\bibfnamefont{M.}~\bibnamefont{Malik}},
  \emph{\bibinfo{title}{High-dimensional pixel entanglement: Efficient
  generation and certification}} (\bibinfo{year}{2020}), \eprint{2004.04994}.

\bibitem[{\citenamefont{Weilenmann et~al.}(2019)\citenamefont{Weilenmann, Dive,
  Trillo, Aguilar, and Navascués}}]{weilenmann2019entanglement}
\bibinfo{author}{\bibfnamefont{M.}~\bibnamefont{Weilenmann}},
  \bibinfo{author}{\bibfnamefont{B.}~\bibnamefont{Dive}},
  \bibinfo{author}{\bibfnamefont{D.}~\bibnamefont{Trillo}},
  \bibinfo{author}{\bibfnamefont{E.~A.} \bibnamefont{Aguilar}},
  \bibnamefont{and}
  \bibinfo{author}{\bibfnamefont{M.}~\bibnamefont{Navascués}},
  \emph{\bibinfo{title}{Entanglement detection beyond measuring fidelities}}
  (\bibinfo{year}{2019}), \eprint{1912.10056}.

\bibitem[{\citenamefont{Hill and Wootters}(1997)}]{Hill_1997}
\bibinfo{author}{\bibfnamefont{S.}~\bibnamefont{Hill}} \bibnamefont{and}
  \bibinfo{author}{\bibfnamefont{W.~K.} \bibnamefont{Wootters}},
  \bibinfo{journal}{Physical Review Letters} \textbf{\bibinfo{volume}{78}},
  \bibinfo{pages}{5022–5025} (\bibinfo{year}{1997}), ISSN
  \bibinfo{issn}{1079-7114},
  \urlprefix\url{http://dx.doi.org/10.1103/PhysRevLett.78.5022}.

\bibitem[{\citenamefont{Schneeloch and Howland}(2018)}]{Schneeloch_2018}
\bibinfo{author}{\bibfnamefont{J.}~\bibnamefont{Schneeloch}} \bibnamefont{and}
  \bibinfo{author}{\bibfnamefont{G.~A.} \bibnamefont{Howland}},
  \bibinfo{journal}{Physical Review A} \textbf{\bibinfo{volume}{97}}
  (\bibinfo{year}{2018}), ISSN \bibinfo{issn}{2469-9934},
  \urlprefix\url{http://dx.doi.org/10.1103/PhysRevA.97.042338}.

\bibitem[{\citenamefont{Schneeloch et~al.}(2019)\citenamefont{Schneeloch,
  Tison, Fanto, Alsing, and Howland}}]{Schneeloch_2019}
\bibinfo{author}{\bibfnamefont{J.}~\bibnamefont{Schneeloch}},
  \bibinfo{author}{\bibfnamefont{C.~C.} \bibnamefont{Tison}},
  \bibinfo{author}{\bibfnamefont{M.~L.} \bibnamefont{Fanto}},
  \bibinfo{author}{\bibfnamefont{P.~M.} \bibnamefont{Alsing}},
  \bibnamefont{and} \bibinfo{author}{\bibfnamefont{G.~A.}
  \bibnamefont{Howland}}, \bibinfo{journal}{Nature Communications}
  \textbf{\bibinfo{volume}{10}} (\bibinfo{year}{2019}), ISSN
  \bibinfo{issn}{2041-1723},
  \urlprefix\url{http://dx.doi.org/10.1038/s41467-019-10810-z}.

\bibitem[{\citenamefont{Berta et~al.}(2010)\citenamefont{Berta, Christandl,
  Colbeck, Renes, and Renner}}]{Berta_2010}
\bibinfo{author}{\bibfnamefont{M.}~\bibnamefont{Berta}},
  \bibinfo{author}{\bibfnamefont{M.}~\bibnamefont{Christandl}},
  \bibinfo{author}{\bibfnamefont{R.}~\bibnamefont{Colbeck}},
  \bibinfo{author}{\bibfnamefont{J.~M.} \bibnamefont{Renes}}, \bibnamefont{and}
  \bibinfo{author}{\bibfnamefont{R.}~\bibnamefont{Renner}},
  \bibinfo{journal}{Nature Physics} \textbf{\bibinfo{volume}{6}},
  \bibinfo{pages}{659–662} (\bibinfo{year}{2010}), ISSN
  \bibinfo{issn}{1745-2481},
  \urlprefix\url{http://dx.doi.org/10.1038/nphys1734}.

\bibitem[{\citenamefont{Carlen and Lieb}(2012)}]{Carlen_2012}
\bibinfo{author}{\bibfnamefont{E.~A.} \bibnamefont{Carlen}} \bibnamefont{and}
  \bibinfo{author}{\bibfnamefont{E.~H.} \bibnamefont{Lieb}},
  \bibinfo{journal}{Letters in Mathematical Physics}
  \textbf{\bibinfo{volume}{101}}, \bibinfo{pages}{1–11}
  (\bibinfo{year}{2012}), ISSN \bibinfo{issn}{1573-0530},
  \urlprefix\url{http://dx.doi.org/10.1007/s11005-012-0565-6}.

\bibitem[{\citenamefont{Coles et~al.}(2017)\citenamefont{Coles, Berta,
  Tomamichel, and Wehner}}]{RevModPhys.89.015002}
\bibinfo{author}{\bibfnamefont{P.~J.} \bibnamefont{Coles}},
  \bibinfo{author}{\bibfnamefont{M.}~\bibnamefont{Berta}},
  \bibinfo{author}{\bibfnamefont{M.}~\bibnamefont{Tomamichel}},
  \bibnamefont{and} \bibinfo{author}{\bibfnamefont{S.}~\bibnamefont{Wehner}},
  \bibinfo{journal}{Rev. Mod. Phys.} \textbf{\bibinfo{volume}{89}},
  \bibinfo{pages}{015002} (\bibinfo{year}{2017}),
  \urlprefix\url{https://link.aps.org/doi/10.1103/RevModPhys.89.015002}.

\end{thebibliography}
\appendix
\begin{widetext}
\section{Measuring individual density matrix elements}

In our scheme of quantifying entanglement, we need to measure the values of $\langle ij|\rho|ij\rangle$ and the real part of the off-diagonal terms $\Re e[\langle ii|\rho|jj\rangle]$ in the density matrix. The diagonal elements are given by $\langle ij |\rho|ij \rangle=C(ij)/C_T$, where $C_T:=\sum_{ij=0}^{31}{C(ij)}$ are the normalised coincidences in path $i$ for Alice and path $j$ for Bob. These diagonal elements can be measured directly.  The off-diagonal terms require a superposition of two paths to be measured on both sides.
\begin{equation}
\Re e[\langle i i|\rho| j j\rangle]=\frac{1}{4}\left(\langle\sigma_{x}^{ij}\otimes\sigma_{x}^{ij}\rangle-\langle\sigma_{y}^{ij}\otimes\sigma_{y}^{ij}\rangle\right),
\end{equation}
\begin{equation}
\Im m[\langle i i|\rho| j j\rangle]=i\frac{1}{4}\left(\langle\sigma_{x}^{ij}\otimes\sigma_{y}^{ij}\rangle-\langle\sigma_{y}^{ij}\otimes\sigma_{x}^{ij}\rangle\right),
\end{equation}
where $\sigma_{ x }^{ ab }=| a \rangle\langle b |+| b \rangle\langle a |$ and $\sigma_{ y }^{ ab }= i | a \rangle\langle b | - i | b \rangle\langle a |$.

This requires projective measurements of any two dimensional subspace. Out of the 992 off-diagonal elements, only 496 are unique and need to be measured, as $\Re e[\langle ii|\rho|jj\rangle]=\Re e[\langle jj|\rho|ii\rangle]$, with $i,j=0,1,\dots,31 (i<j)$.  The average interference visibility of all subspaces is $V=0.974\pm0.001$, leading to the following fidelities:

\begin{table*}[btph]
\caption{Experimental results of fidelity in different dimension.}
\begin{tabular}{c|c|c|c|c}
  \toprule
  \hline
  Dimension & 2 & 4 & 6 & 8  \\
  \hline
  Fidelity(\%) & $98.8\pm0.1$ & $97.9\pm0.1$ & $97.6\pm0.1$ & $97.6\pm0.1$   \\
  \hline
  Dimension & 10 & 12 & 14 & 16  \\
  \hline
  Fidelity(\%) & $97.3\pm0.1$ & $97.1\pm0.1$ & $96.6\pm0.1$ & $96.3\pm0.1$   \\
    \hline
  Dimension & 18 & 20 & 22 & 24  \\
  \hline
  Fidelity(\%) & $96.1\pm0.1$ & $95.8\pm0.1$ & $95.7\pm0.1$ & $95.3\pm0.1$   \\
    \hline
  Dimension & 26 & 28 & 30 & 32  \\
  \hline
  Fidelity(\%) & $94.8\pm0.1$ & $94.3\pm0.1$ & $93.8\pm0.1$ & $93.3\pm0.1$ \\
\end{tabular}
\label{table:fidelity}
\end{table*}

\section{Bounding Entanglement of Formation via Entropic Uncertainty Relations}

Given the measurement data of the correlations in the computational basis as well as the respective coherences, i.e.
\begin{align}
\langle ii | \rho | jj  \rangle ,
\end{align}
and thus the fidelity with the maximally entangled state $F_{+}:=F (\rho, |\phi^{+}\rangle\langle\phi^{+} |)$, we can derive a lower bound on entanglement of formation.

Using the uncertainty relation derived in \cite{Berta_2010} with extensions presented in \cite{RevModPhys.89.015002},
\begin{align}
H(M|B) + H(\tilde{M}|B) \geq H(A|B) + log(d)\,,
\end{align}
where $M$ and $\tilde{M}$ are two mutually unbiased bases in system A and $H(M/\tilde{M}|B)$ is the conditional von Neumann entropy of the state after measurement of $M/\tilde{M}$ in system A. We isolate the term $H(A|B)$, as it lower bounds the Entanglement of Formation \cite{Carlen_2012}

\begin{align}
- H(A|B) \leq E_{oF}.
\end{align}

This leads to

\begin{align}
E_{oF} \geq - H(M|B) - H(\tilde{M}|B) + log(d).
\end{align}

Expanding the relation and using the data processing inequality \cite{RevModPhys.89.015002} as well as the definition of the relative entropy we arrive at

\begin{align}
E_{oF} \geq  - H(M,M^*) - H(\tilde{M},\tilde{M}^*) +  H(M)+  H(\tilde{M}) + log(d)\,.
\end{align}

We can further lower bound this expression by upper bounding the global entropies with $H^{\uparrow}$ and lower bounding the marginal entropy by $H^{\downarrow}$,
\begin{align}
E_{oF} \geq  - H^{\uparrow}(M,M^*) - H^{\uparrow}(\tilde{M},\tilde{M}^*) +  H^{\downarrow}(M)+  H^{\downarrow}(\tilde{M}) + log(d)\,.
\end{align}

Now we just need suitable bounds $H^{\uparrow/\downarrow}$ for our measurement data. Let us first discuss the bounds on the entropies for the measurement results in the computational basis with the help of the given quantities. It will be helpful to consider the following abbreviations
\begin{align}
&p^{AB}_i :=   \langle i i | \rho | i i \rangle \\
&N := \sum_{i}  \langle i i | \rho | i i \rangle,
\end{align}
the marginal probabilities can be directly lower bounded by using
\begin{align}
\langle i_{B} | \rho | i_{B} \rangle &= \sum_{j}   \langle j i | \rho | j i \rangle \\
&=  \langle i i | \rho | i i \rangle + \sum_{j\neq i}   \langle j i | \rho | j i \rangle \\
&\geq \langle i i | \rho | i i \rangle  := p^{B}_i.
\end{align}
Now the next task is to minimise the entropy, given these lower bounds, which is straightforward: The entropy is  minimised by adding $1-N$ to the largest $p^{B}_i$, creating a distribution that majorises all other valid marginal  distributions and thus possessing the lowest entropy. After this transformation, we just need to compute the entropy of that distribution
\begin{align}
 H^{\downarrow}(M) = - \sum_{i} p^{B}_i log (p^{B}_i ).
\end{align}
For the upper bound on the global entropy on the other hand we can use
\begin{align}
H^{\uparrow}(M,M) = - \sum_{i} p^{AB}_i log(p^{AB}_i  ) -  \left[ (1-N) log \left( (1-N)\frac{1}{\left( d^2 - d \right)} \right)\right].
\end{align}

For deriving the bounds on the entropies for the measurement results in the MUB, we'll use the invariance of the target state under rotations of the form $U\otimes U^*$. This lets us conclude that
\begin{align}
    \langle \tilde{i}\tilde{i}^*|\rho|\tilde{i}\tilde{i}^*\rangle=\frac{F_+}{d}+\tilde{p}_i\,,
\end{align}
where $\tilde{p}_i\geq 0$ and $\sum_i\tilde{p}_i=1-\frac{(d-1)F_+}{d}$. Thus,
\begin{align}
    H^{\downarrow}(\tilde{M})=H_2\left(\frac{(d-1)F_+}{d}\right)+\frac{(d-1)F_+}{d}\text{log}(d-1)\,,
\end{align}
 where we have used the binary entropy $H_2(p):=-p\text{log}(p)-(1-p)\text{log}(1-p)$.
 
For the upper bound on the global entropy we again use our knowledge of the fidelity $F_{+}$ and equally distribute the remaining norm among the unknown elements, s.t.
\begin{align}
H^{\uparrow}(\tilde{M},\tilde{M}^*) =-\left[F_{+}log\left( \frac{F_{+}}{d} \right) \right]-\left[ (1-F_{+})log\left(\frac{1-F_{+}}{d^2 - d}\right) \right]\,,
\end{align}
which maximises entropy under the assumption that $\frac{F_{+}}{d}\geq\frac{1-F_{+}}{d^2 - d}$.
Using the experimental data in $32$ dimensions, we now get $F_{+}=0.933$, $H^{\downarrow}(M)=4.967$, $H^{\downarrow}(\tilde{M})=4.935$, $H^{\uparrow}(M,M)=5.483$ and $H^{\uparrow}(\tilde{M},\tilde{M}^*)=5.670$. This leads to a lower bound on entanglement of formation of $E_{oF}\geq3.728\pm0.006$. All results are shown in Table~\ref{table:capacity}.

\begin{table*}[btph]
\caption{Experimental results for "e-bits".}
\begin{tabular}{c|c|c|c|c}
   \hline
  Dimension & 2 & 4 & 6 & 8  \\
  \hline
   "e-bit" & $0.380\pm0.006$ & $1.200\pm0.009$ & $1.699\pm0.009$ & $2.056\pm0.009$   \\
  \hline
  Dimension & 10 & 12 & 14 & 16  \\
  \hline
   "e-bit" & $2.333\pm0.009$ & $2.560\pm0.009$ & $2.750\pm0.008$ & $2.914\pm0.009$   \\
    \hline
  Dimension & 18 & 20 & 22 & 24  \\
  \hline
   "e-bit" & $3.055\pm0.007$ & $3.185\pm0.007$ & $3.304\pm0.007$ & $3.407\pm0.006$   \\
    \hline
  Dimension & 26 & 28 & 30 & 32  \\
  \hline
   "e-bit" & $3.499\pm0.006$ & $3.583\pm0.006$ & $3.660\pm0.006$ & $3.728\pm0.006$ \\
\end{tabular}
\label{table:capacity}
\end{table*}

\section{Noise robustness}

Whether the theoretical robustness to noise of high-dimensionally entangled states can be reached in practice depends on various parameters. Most importantly, the quality of the source as well as the nature of the noise. A standard method, that e.g. models accidentals for photon loss of maximally entangled states leads to white noise, i.e. the state actual state will resemble $\rho_{exp}=p\rho_{target}+\frac{1-p}{d^2}\mathbbm{1}$. While the actual dependence of $p$ on the dimension will heavily depend on the implementation parameters, it is nonetheless crucial for $\rho_{target}$, to be as close to $|\Phi^+\rangle\langle\Phi^+|$ in laboratory situations. This is why we focus on determining $F_{+}$. For there to be a possible advantage in noise resistance, this fidelity needs to scale better than ${F}_{sep}=\frac{1}{d}$. This means our experimental fidelities certify that every dimension we add, brings about meaningful additional noise robustness, despite declining fidelity. 

\section{Construction of complete MUB measurement setup}

Some Bell inequalities and quantum key distribution (QKD) protocols based on entanglement require complete MUB measurements. In this section, we will introduce how to construct a complete MUB measurement scheme with our setup. The most typical MUB is the computation basis ($\{|\ell_{0}\rangle=|0\rangle, \ldots,|\ell_{k}\rangle=|k\rangle,\ldots|\ell_{d-1}\rangle=|d-1\rangle\}$) and the Fourier basis ($\{\left|L_{0}\right\rangle, \ldots,\left|L_{k}\right\rangle, \ldots,\left|L_{d-1}\right\rangle\}$). $\left|L_{k}\right\rangle$ is the superposition of $\left|\ell_{k}\right\rangle, \left|L_{k}\right\rangle=\sum_{j=0}^{d-1} \exp [(i 2 \pi / d) k j]\left|\ell_{j}\right\rangle / \sqrt{d}$. Here, we introduce how to construct $2^{n}$-dimensional computational and Fourier MUB basis. Fourier MUB basis are coherent superposition of all levels with equal weight. The simplest basis construction in our context is the direct use of product basis of qubit mutually unbiased bases to construct a set of mutually orthogonal high-dimensional MUB basis. The simplest set of MUB bases for qubit system is $\{1/\sqrt{2}(|0\rangle+|1\rangle), 1/\sqrt{2}(|0\rangle-|1\rangle)\}$. If we take a direct product of multiple MUB bases of $n$ qubits $\{1/\sqrt{2}(|0\rangle\pm|1\rangle)_{1}\otimes 1/\sqrt{2}(|0\rangle\pm|1\rangle)_{2}\otimes, ..., \otimes 1/\sqrt{2}(|0\rangle\pm|1\rangle)_{n} \}$, and then encode $\{|00...0\rangle, |00...1\rangle,..., |11...1\rangle \}$ into $\{|0\rangle, |1\rangle,..., |2^{d}-1\rangle \}$, we can get a set of MUB bases of $d=2^{n}$ dimensions. In Fig.~\ref{fig:mub}, we give the scheme of $d=4, d=8$, and any $d=2^{n}$ dimensions. We take the 8-dimensional MUB construction base as an example. The 8-dimensional MUB can be obtained from the MUB direct product of three qubits ($\{1/\sqrt{2}(|0\rangle\pm|1\rangle)_{1}\otimes 1/\sqrt{2}(|0\rangle\pm|1\rangle)_{2}\otimes 1/\sqrt{2}(|0\rangle\pm|1\rangle)_{3} \}$). We code the three qubit system into a 8-dimensional system ($\{|000\rangle, |001\rangle, |010\rangle, |011\rangle, |100\rangle, |101\rangle, |110\rangle,|111\rangle \}$ into $\{|0\rangle, |1\rangle, |2\rangle, |3\rangle, |4\rangle, |5\rangle, |6\rangle,|7\rangle \}$). HWPA1 and BD1 are used to merge adjacent dimensions into two-dimensional polarization subspace. HWP1 and PBS1 construct the MUB of the two-dimensional subspace, which is equivalent to the construction of the MUB base of the third qubit. Similarly, HWPA2, BD2, HWP2 and PBS2 complete the MUB construction of the second qubit, while HWPA3, BD3, HWP3 and PBS3 complete the MUB construction of the first qubit. By cascading these three setups and eight outcomes, an 8-dimensional complete MUB measurement base is constructed. All HWPs are set at $22.5^{o}$. SLM1 is used to load different phases on different paths, so that different Fourier MUB bases can be converted. For the computational MUB basis, we still take the 8-dimensional measurement in Fig.~\ref{fig:mub}b. We only need to set the HWP1-3 in Fig.~\ref{fig:mub}b to $0^{o}$ to realize the 8-dimensional computational measurement basis. As shown in Fig.~\ref{fig:mub}c, this method can be effectively extended to any $d = 2^n$ dimension that is a power of $2$, which are anyway convenient for encoding multiple bits in a photon. 

SLM1 and HWP1-3 can be replaced by fast electro-optical modulation setups, so the MUB measurement setup designed by us could also be used for a high-speed high-dimensional QKD device, or the detection of Bell inequality without an independent measurement loophole.

\begin{figure*}[tbph]
\includegraphics [width= 1\textwidth]{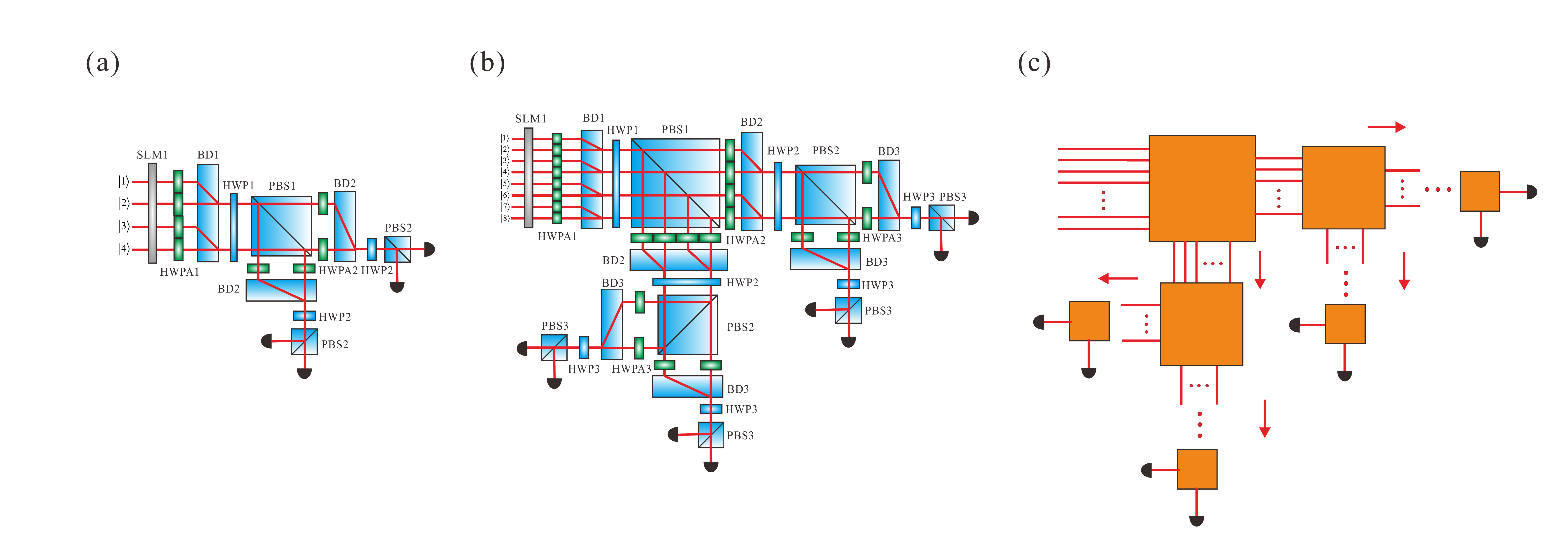}
\caption{Construction of $d=2^{n}$ dimensional MUB measurement basis. (a) 4-dimensional MUB measurement structure. (b) 8-dimensional MUB measurement structure. (c) $2^{n}$-dimensional MUB measurement structure.  }
\label{fig:mub}
\end{figure*}

\section{Measurement basis settings}

Here we present some HWPA settings for 2-dimensional subspace projection measurements. In fact, different projection measurements are realised by changing the settings of HWPAs or HWPs. These devices can be replaced by adjustable phase devices (such as SLMs), so they do not need to be replaced to achieve full-automatic measurement.

\begin{table*}[btph]
\caption{$|0i\rangle$ subspaces settings. SSM represents the wave plates projected to the subspace for measurement control. By controlling these wave plates, we can realise arbitrary projection measurements in this two-dimensional subspace. }
\begin{tabular}{c|c|c|c|c|c}
  \toprule
  \hline
   & HWPA2 & HWPA3 & HWPA4 & HWPA5 & HWP \\
  \hline
   (0,1) & SSM & $HWP@0^{\circ}$ & $HWP@0^{\circ}$ & $HWP@0^{\circ}$ & $HWP1@45^{\circ}$\\ \hline
   (0,2)  & $\theta 2@0^{\circ}$  & SSM    & $HWP@0^{\circ}$ & $HWP@0^{\circ}$ & $HWP1@45^{\circ}$ \\
   \hline
(0,3)  & $HWP@0^{\circ}$ & SSM    & $HWP@0^{\circ}$ & $HWP@0^{\circ}$ & $HWP1@45^{\circ}$ \\
\hline
(0,4)  & $\theta 2@0^{\circ}$  & $\theta 3@0^{\circ}$  & SSM    & $HWP@0^{\circ}$ & $HWP1@45^{\circ}$ \\
\hline
(0,5)  & $\theta 3@0^{\circ}$  & $\theta 3@0^{\circ}$  & SSM    & $HWP@0^{\circ}$ & $HWP1@45^{\circ}$ \\
\hline
(0,6)  & $\theta 2@0^{\circ}$  & $HWP@0^{\circ}$ & SSM    & $HWP@0^{\circ}$ & $HWP1@45^{\circ}$ \\
\hline
(0,7)  & $HWP@0^{\circ}$ & $HWP@0^{\circ}$ & SSM    & $HWP@0^{\circ}$ & $HWP1@45^{\circ}$ \\ \hline
(0,8)  & $HWP@0^{\circ}$ & $HWP@0^{\circ}$ & $\theta 2@90^{\circ}$ & SSM    & $HWP1@45^{\circ}$ \\
\hline
(0,9)  & $\theta 2@90^{\circ}$ & $HWP@0^{\circ}$ & $\theta 2@90^{\circ}$ & SSM    & $HWP1@45^{\circ}$ \\
\hline
(0,10) & $\theta 2@0^{\circ}$  & $\theta 2@90^{\circ}$ & $\theta 2@90^{\circ}$ & SSM    & $HWP1@45^{\circ}$ \\
\hline
(0,11) & $HWP@0^{\circ}$ & $\theta 2@90^{\circ}$ & $\theta 2@90^{\circ}$ & SSM    & $HWP1@45^{\circ}$ \\
\hline
(0,12) & $HWP@0^{\circ}$ & $\theta 3@0^{\circ}$  & $HWP@0^{\circ}$ & SSM    & $HWP1@45^{\circ}$ \\
\hline
(0,13) & $\theta 3@0^{\circ}$  & $\theta 3@0^{\circ}$  & $HWP@0^{\circ}$ & SSM    & $HWP1@45^{\circ}$ \\
\hline
(0,14) & $\theta 3@0^{\circ}$  & $HWP@0^{\circ}$ & $HWP@0^{\circ}$ & SSM    & $HWP1@45^{\circ}$ \\
\hline
(0,15) & $HWP@0^{\circ}$ & $HWP@0^{\circ}$ & $HWP@0^{\circ}$ & SSM    & $HWP1@45^{\circ}$ \\ \hline
(0,16) & $HWP@0^{\circ}$ & $HWP@0^{\circ}$ & $HWP@0^{\circ}$ & $\theta 3@90^{\circ}$ & SSM      \\
\hline
(0,17) & $\theta 3@90^{\circ}$ & $HWP@0^{\circ}$ & $HWP@0^{\circ}$ & $\theta 3@90^{\circ}$ & SSM      \\
\hline
(0,18) & $\theta 2@0^{\circ}$  & $\theta 3@90^{\circ}$ & $HWP@0^{\circ}$ & $\theta 3@90^{\circ}$ & SSM      \\
\hline
(0,19) & $HWP@0^{\circ}$ & $\theta 3@90^{\circ}$ & $HWP@0^{\circ}$ & $\theta 3@90^{\circ}$ & SSM      \\
\hline
(0,20) & $HWP@0^{\circ}$ & $\theta 3@0^{\circ}$  & $\theta 3@90^{\circ}$ & $\theta 3@90^{\circ}$ & SSM      \\
\hline
(0,21) & $\theta 3@0^{\circ}$  & $\theta 3@0^{\circ}$  & $\theta 3@90^{\circ}$ & $\theta 3@90^{\circ}$ & SSM      \\
\hline
(0,22) & $\theta 3@0^{\circ}$  & $HWP@0^{\circ}$ & $\theta 3@90^{\circ}$ & $\theta 3@90^{\circ}$ & SSM      \\
\hline
(0,23) & $HWP@0^{\circ}$ & $HWP@0^{\circ}$ & $\theta 3@90^{\circ}$ & $\theta 3@90^{\circ}$ & SSM      \\
\hline
(0,24) & $HWP@0^{\circ}$ & $HWP@0^{\circ}$ & $\theta 3@90^{\circ}$ & $HWP@0^{\circ}$ & SSM      \\
\hline
(0,25) & $\theta 3@90^{\circ}$ & $HWP@0^{\circ}$ & $\theta 3@90^{\circ}$ & $HWP@0^{\circ}$ & SSM      \\
\hline
(0,26) & $\theta 2@0^{\circ}$  & $\theta 3@90^{\circ}$ & $\theta 3@90^{\circ}$ & $HWP@0^{\circ}$ & SSM      \\
\hline
(0,27) & $HWP@0^{\circ}$ & $\theta 3@90^{\circ}$ & $\theta 3@90^{\circ}$ & $HWP@0^{\circ}$ & SSM      \\
\hline
(0,28) & $HWP@0^{\circ}$ & $\theta 3@0^{\circ}$  & $HWP@0^{\circ}$ & $HWP@0^{\circ}$ & SSM      \\
\hline
(0,29) & $\theta 3@0^{\circ}$  & $\theta 3@0^{\circ}$  & $HWP@0^{\circ}$ & $HWP@0^{\circ}$ & SSM      \\
\hline
(0,30) & $\theta 3@0^{\circ}$  & $HWP@0^{\circ}$ & $HWP@0^{\circ}$ & $HWP@0^{\circ}$ & SSM      \\
\hline
(0,31) & $HWP@0^{\circ}$ & $HWP@0^{\circ}$ & $HWP@0^{\circ}$ & $HWP@0^{\circ}$ & SSM      \\ \hline
\end{tabular}
\label{table:angles}
\end{table*}
\end{widetext}
\end{document}